\documentclass[myepj-spec,pdftex]{mySvjour}
\usepackage{graphicx}
\usepackage[pdfpagemode=UseNone]{hyperref}
\usepackage{color}
\usepackage{xspace}
\usepackage[square,sort&compress,numbers]{natbib}   
\usepackage{verbatim}
\usepackage{amsmath,amssymb,amsfonts} 
\usepackage{tabularx}
\usepackage{multicol}
\usepackage{footnote}
\usepackage{listings}
\usepackage[gen]{eurosym}
\usepackage[switch, modulo]{lineno}
\usepackage{lmodern,bm}                
\usepackage[T1]{sansmath} 
\SetMathAlphabet{\mathsfbf}{sans}{\sansmathencoding}{\sfdefault}{bx}{sl}
\usepackage{etoolbox}
\AtBeginEnvironment{sansmath}{}{}{}

\usepackage{lipsum}

\newcommand{\June}{\textsc{June}\xspace}
\newcommand{\JuneGermany}{\textsc{June-Germany}\xspace}

\definecolor{darkblue1}{rgb}{0,0,.2}
\definecolor{darkblue}{rgb}{0,0,.2}
\definecolor{darkred}{rgb}{0.5,0,0}
\pagecolor{white} 
\color{black}     
\hypersetup{breaklinks=true, 
	colorlinks=true, 
	linkcolor=darkblue1, 
	menucolor=darkblue1, 
	urlcolor=darkblue1,
	citecolor=darkblue1,
	pdftitle={},
	pdfauthor={},
	pdfsubject={},
	pdfkeywords={},
	pdfproducer={}
}
\usepackage{siunitx}

\bibstyle{plain}

\begin{document}
	
	\twocolumn[{%
		\begin{@twocolumnfalse}
			
			\begin{flushright}
				\normalsize
			\end{flushright}
			
			\vspace{-2cm}
			
			\title{\Large\boldmath Why a Mayor cannot Change the Course of a Pandemic - An agent-based Study on the Covid Spread on Local Level in Germany}
			%

\author{Lucas Heger\inst{1}, Kerem Akdogan\inst{1} \and Matthias Schott \inst{1,2}}
\institute{%
    \inst{1} Institute of Physics, Johannes Gutenberg University, Mainz, Germany\\
    \inst{2} Corresponding author: schottm@uni-mainz.de
}

			
			\abstract{During the COVID-19 pandemic, a large variance of incidence rates on local level, e.g. cities and districts, within one country has been observed, while the same non-pharmaceutical measures have been taken to control the spread of the virus. This variance in incidence rates triggered the question, if the spread of incidence rates can be explained only by statistical processes and the local population statistics or if indeed other factors, e.g. local information campaigns, have to be considered. Within this paper we study the expected spread of incidence rates in the German State of Rhineland Palatinate during the second COVID-19 wave using an agent based simulation and find that the spread of incidence rates can be solely explained by population statistics and further statical effects.} 
	\maketitle
	\end{@twocolumnfalse}
}]


\section{Introduction}

Amidst the global COVID-19 pandemic, nations took various measures to curb the virus's spread. These encompassed extensive lockdowns, movement restrictions, social distancing guidelines, quarantine procedures, mask mandates, and robust testing and contact tracing initiatives. Travel restrictions were enforced, while remote work and online education were promoted. Most measures have been taken at state- or national wide level, however, still a significant variance on the actual infection rates on local level, e.g. districts or cities has been observed. This triggered the assumption, mainly in media, that some kind of special local circumstances, e.g. the impact of local politics, testing- or information campaigns might cause this variance \cite{Media1, Media2}. 

Within this study we evaluate if the observed spread in incidence rates can be explained solely by population related statistics or if indeed special local conditions need to be taken into account. We have chosen the state of Rhineland Palatinate as a testing environment for this hypothesis. Given that Rhineland Palatinate is a medium sized state within Germany with 3.6 million inhabitants, including larger cities and more rural areas, we would argue that the general conclusion should stay largely valid for the whole of Germany. 

Mathematical simulations and forecasts for epidemic or pandemic evolution commonly rely on compartmental models or agent-based models. Compartmental models categorise the population into compartments (e.g., susceptible, exposed, infected, recovered) and employ differential equations to depict transitions between these states (e.g., SIR model \cite{articleWangSir, articleSir2}). In contrast, agent-based models simulate individual agents with specific characteristics and behaviours, providing a more intricate representation of interactions and spatial dynamics \cite{Bullock:2021, pillai2023agentbased, muller2020realistic, articleABM1}. We chose the latter modelling approach for our study, given that agent based models can describe detailed population statistics and the interaction between agents as well as their commuting behaviour. 

The agent-based simulation, which is used for our study is briefly described in Section \ref{sec:framework}, followed by a summary of the predictive power of the framework within RLP during the second Covid-19 wave between October 2020 and February 2021 (Section \ref{sec:simRLP}). The observed statistical spread of infection incidences is discussed and interpreted in Section \ref{sec:results}, followed by a brief conclusion.

\begin{figure*}[thb]
\centering
    \includegraphics[width=0.49\textwidth]{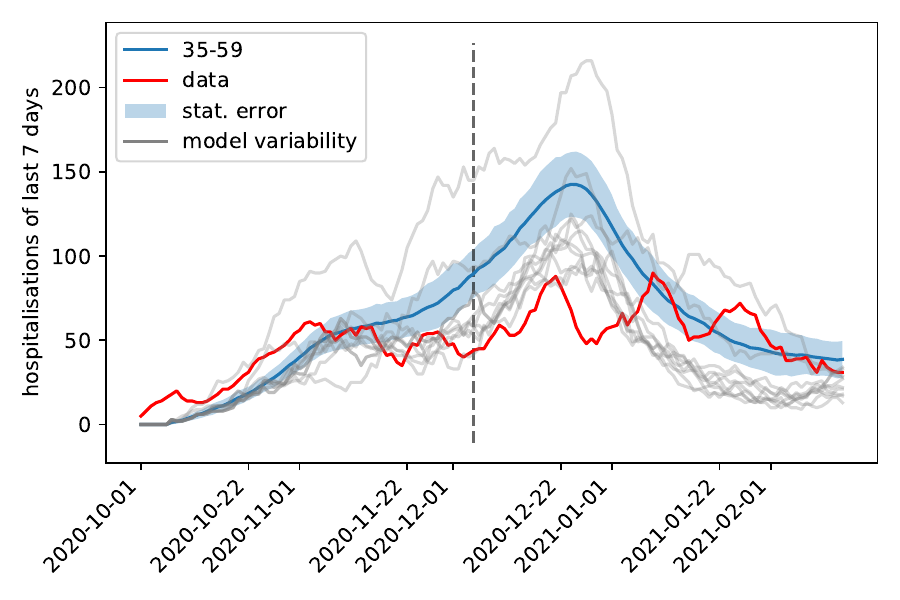}
    \includegraphics[width=0.49\textwidth]{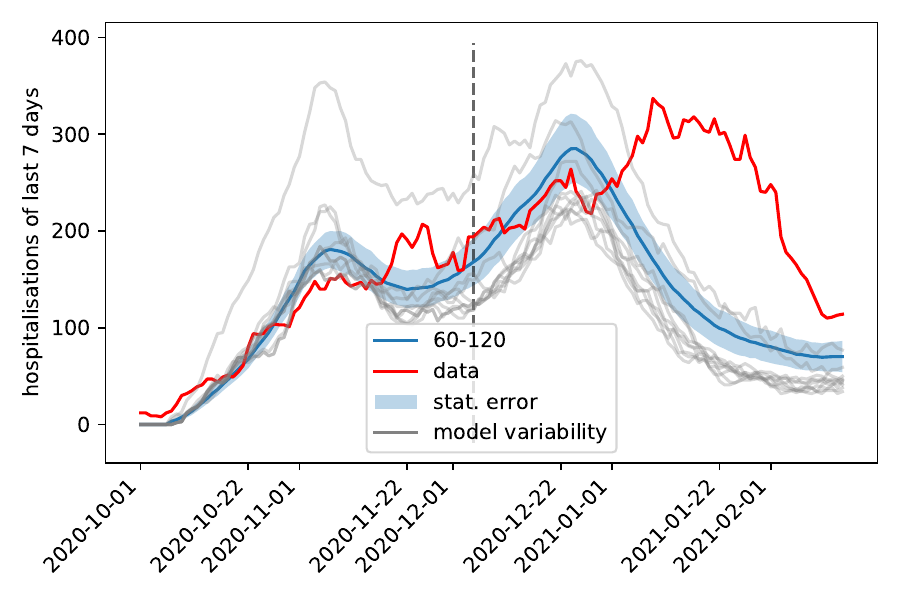}
    \caption{Hospitalization rate of the last 7 days  (red line) for the age group 35-59 (left) and $\geq$60 (right), together with the \JuneGermany simulation with the best fitting parameters (blue line), its statistical uncertainty (shaded blue) and the simulated curves of alternative sets of model parameters tested during the fitting procedure (gray). The vertical line indicates the date until when the data was fitted.}
\label{fig:hos1}
\end{figure*}

\section{The \JuneGermany Framework\label{sec:framework}}

The \June framework, introduced by Bullock et al. in 2021 \cite{Bullock:2021}, is an agent-based model designed to simulate epidemics in a population, specifically focusing on the initial and subsequent waves of the Covid-19 pandemic. Notably, this model incorporates detailed geographic and sociological data for England. \June has demonstrated its ability to accurately forecast the geographical and sociological dynamics of Covid-19 transmission, as discussed in detail in the original publication \cite{Bullock:2021}.

Building upon the success of \June, \JuneGermany \cite{junegermany} adapts the framework for application in Germany. Like its predecessor, it is implemented in \textsc{Python} and structured into four interconnected layers: 

\begin{itemize}
\item {\bf Population Layer:} This layer details individual agents, their static social environments (e.g., households, workplaces), and demographics across hierarchical geographic layers. Agents follow daily routines in discrete time-steps, associated with specific households, schools, and workplaces.

\item {\bf Interaction Layer:} Captures daily routines such as commuting and leisure activities. Social contact networks define interactions, and disease transmission during public transportation is considered. Age-dependent social interaction matrices model contact frequency and intensity in various settings.

\item {\bf Disease Layer:} Models disease transmission and effects, utilising probabilistic infection modelling that considers factors like transmissive probability, susceptibility, and exposure time. Health impacts range from asymptomatic cases to ICU admission and potentially lethal outcomes.

\item {\bf Policy Layer:} Incorporates government policies for pandemic mitigation at localised levels, considering geographical regions and social interactions. This allows for the modelling of essential workers' activities and general population compliance, with agent compliance influenced by social and demographic parameters.

\end{itemize}

The geographical model is based on German administrative areas, featuring three layers: states, districts, and municipalities, with detailed demographic data from the 2011 Census \cite{Zensus2011}. Population densities vary based on age and sex distribution, considering age as a significant risk factor for severe Covid-19 cases.

Household compositions are also derived from the 2011 Census data, categorised by the number of adults and children in each household. The simulation includes 14,502 primary schools (average 204 students per school) and 13,068 secondary schools (average 506 students per school). A teacher-to-student ratio of 0.12 and class sizes between 20 and 30 students are assumed. Agents are distributed to simulated schools based on their home addresses.

Jobs are classified by sector using the International Standard Industrial Classification, with companies modelled in each district based on the average number of employees for a sector. Workplace assignments during population generation consider mobility data.

Social activities and interactions are modelled similarly in both \June and \JuneGermany. Agents' weekday routines involve work/school, shopping, leisure, and staying at home. Beyond working hours, social activities include visits to cinemas, theatres, pubs, and restaurants, contingent on current state regulations and individual compliance. Commuting is represented by a directed network graph, accounting for both short-distance and long-distance travels.

\section{Simulation of the COVID-19 Pandemic in the State of Rhineland Palatinate\label{sec:simRLP}}

We utilised \JuneGermany to model the second wave of the Covid-19 Pandemic in the German state of Rhine-land-Palatinate spanning from October 2020 to February 2021. The cumulative deaths in the age groups 0-4, 5-14, 15-34, 35-59, and $\geq$60 from October 1, 2020, to December 14, 2021, served as the basis for determining optimal model parameters. Of particular significance were the cumulative deaths in the age group above 35, as minimal deaths were reported for the younger population.

\begin{figure*}[thb]
\centering
    \includegraphics[width=0.9\textwidth]{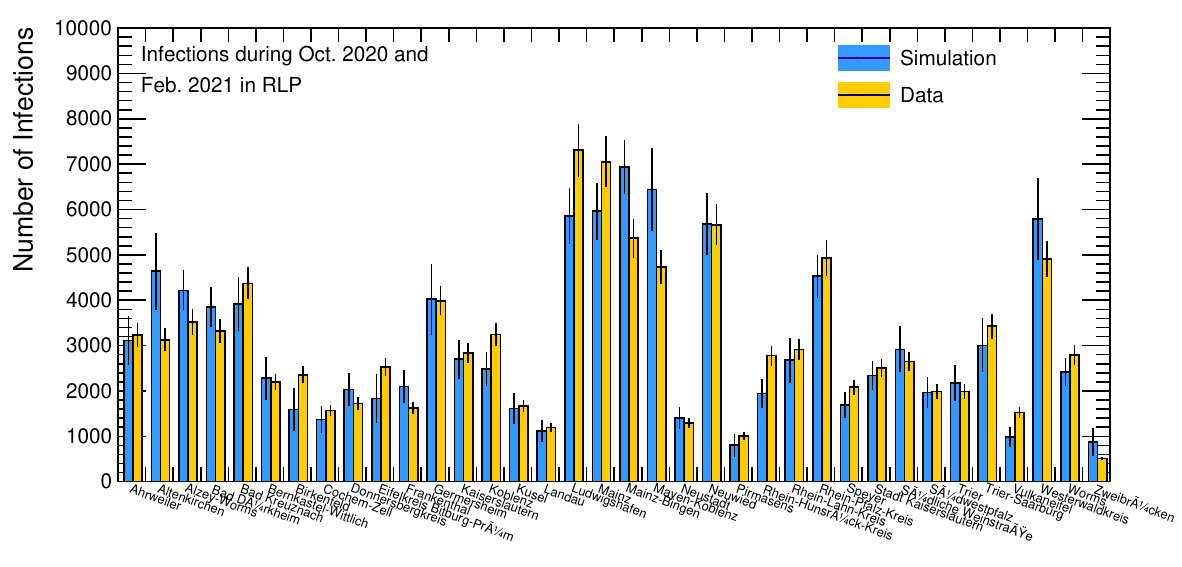}
    \caption{Overall number of infections for all 36 districts of Rhineland Palatinate during October 2020 and February 2021, once for the official reported cases and once by the \JuneGermany simulation.}
\label{fig:res1}
\end{figure*}

\begin{figure*}[thb]
\centering
    \includegraphics[width=0.9\textwidth]{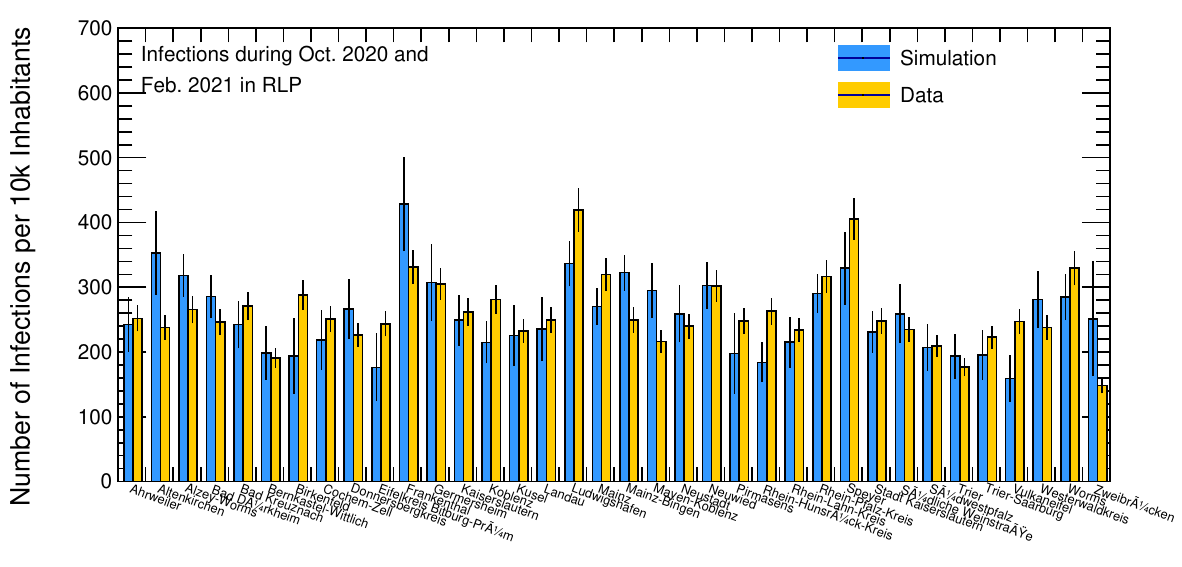}
    \caption{Normalized number of infections per 10k inhabitants for all 36 districts of Rhineland Palatinate during October 2020 and February 2021, once for the official reported cases and once by the \JuneGermany simulation.}
\label{fig:res2}
\end{figure*}

Given the intricacy of the JUNE simulator, characterised by a substantial dimension in both input parameters and output space, we employed emulation and history matching \cite{hme1, hme2} to identify suitable matches to the data. To facilitate this process, we utilised the in-development \textsc{R} package \textsc{hmer} \cite{hmer}, designed to streamline emulator construction and the generation of representative parameter sets for subsequent iterations of emulation and history matching. This package has been employed successfully for parameter estimation in other epidemiological scenarios. The emulation and history matching framework presents several advantages over traditional parameter estimation methods, with a notable benefit being the requirement of relatively few evaluations from the computationally expensive simulator to train an emulator. Following the optimisation of model parameters using data up to December 14, 2021, we projected the entire second wave until February 22, 2022.

The comparison of the hospitalisation rate for the last 7 days is depicted in Figure \ref{fig:hos1}, revealing good agreement, albeit with the simulation predicting a faster decline of the wave than observed in the data. The total number of predicted hospitalised patients during the entire second wave is 5181, compared to the official number of 5638. Notably, a less precise prediction is evident for the incidence rate. Although the number of infections is accurately described for the age group $\geq$60, discrepancies by a factor of about three emerge for the age groups 14-34 and 35-59, indicating a number of unreported cases. This ratio increases significantly for the 5-14 age group. While a general trend toward unreported cases is anticipated, the extent of the observed underreporting appears surprising. A comprehensive discussion can be found in \cite{junegermany}.

\begin{figure*}[h!]
\centering
    \includegraphics[width=0.49\textwidth]{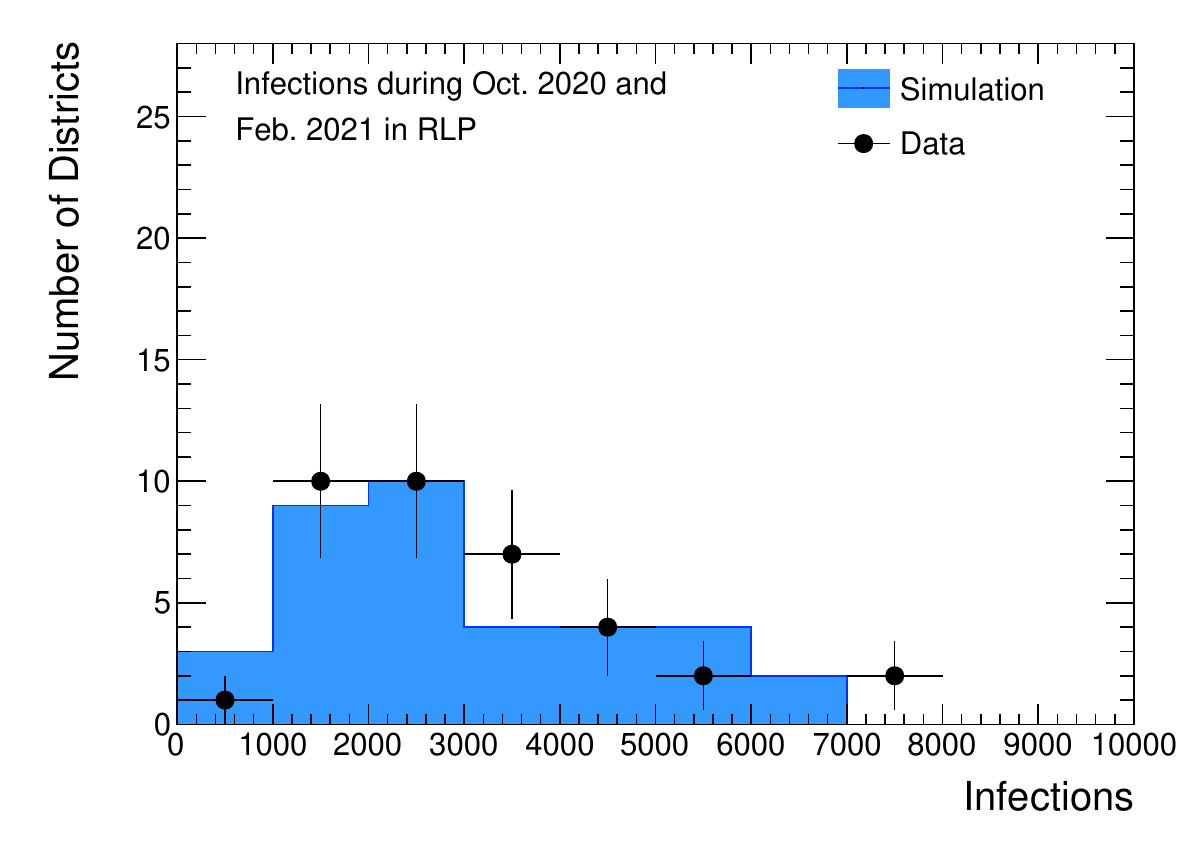}
    \includegraphics[width=0.49\textwidth]{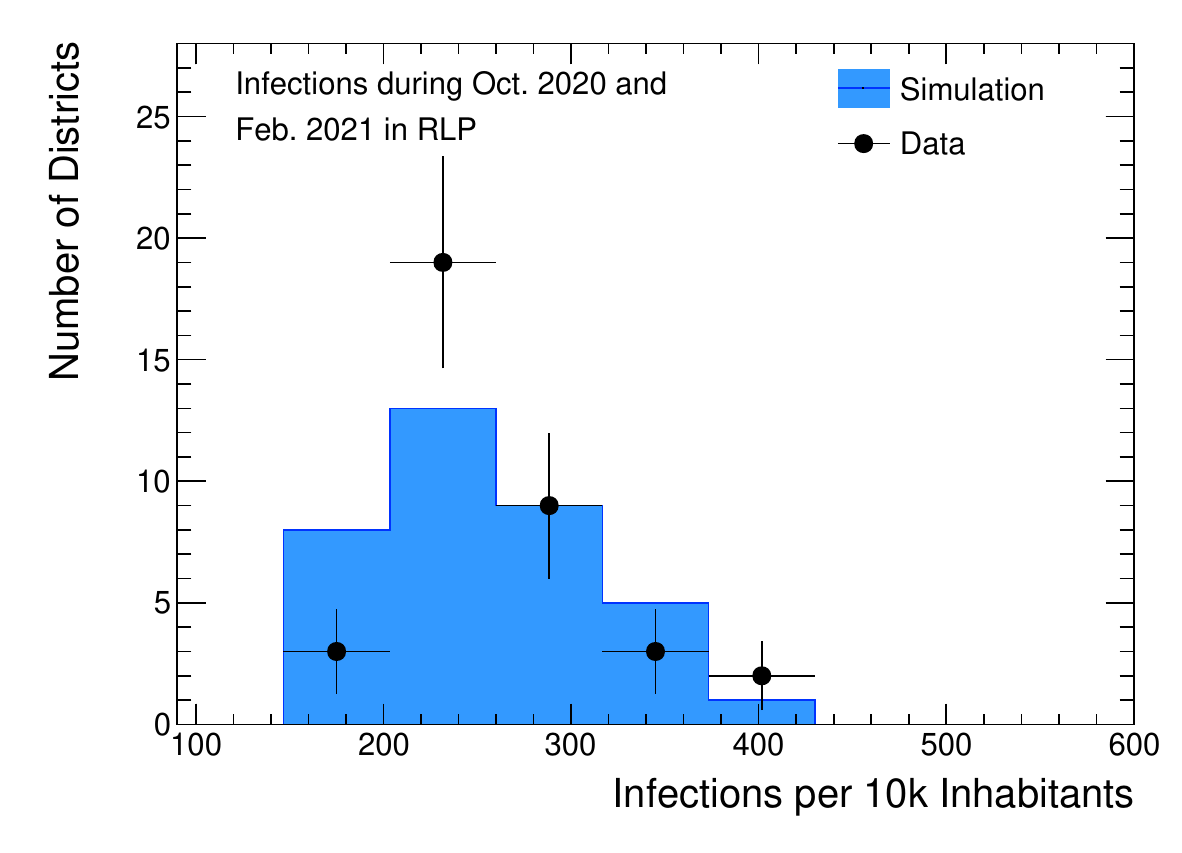}
    \caption{Spread of the overall number of infections (left) and spread of the number of infections per 10k inhabitants in all 36 districts of Rhineland Palatinate during October 2020 and February 2021, once for the official reported cases and once by the \JuneGermany simulation.}
\label{fig:res3}
\end{figure*}

Rhineland Palatinate is structured in 36 districts (Landkreis) and cities. The smallest district is Zwei-brücken with 34k inhabitants, the largest is the city of Mainz as capital with 220k inhabitants. It is important to note, that the optimisation of model parameters is based solely on the overall cumulative deaths rate over all districts and cities and no separate local information has been used. The resulting predictions on the variance among the 36 districts is therefore unbiased.

\section{\label{sec:results} Modelling of Regional Spreads}

Given the observed mismatch between the reported incidence rate and the simulated incidence rate, we first scale the simulated incidence rate to the observed rate with the same global factor across the full population statistics and districts. The relative spread in incidence rates is therefore not affected, however, it simplifies the comparison to the reported data. 

The overall number of reported infections between October 2020 and February 2021 in each of the 36 districts and cities is shown in Figure \ref{fig:res1}. An uncertainty of 5\% on the reported numbers is assumed. Also shown are the simulated numbers of \JuneGermany, where the indicated uncertainties correspond to the 68\% spread of simulation runs with gaussian varied model parameters as well as varied initial conditions to asses statistical uncertainties. A good agreement between simulation and data is observed. As expected, districts and cities with larger population have more cases than those with smaller populations. Hence it is more instructive to compare the number of infections normalised to 10k inhabitants in each district or city, as shown in Figure \ref{fig:res2}. Also here, a very good agreement can be seen. The p-value that reported and simulated numbers agree is 0.2 and rises to 0.4 if the uncertainty on the reported cases per district is 10\%. 

The distribution of the overall numbers infections as well as the normalized number of infections for the observed cases as well as the nominal simulation is shown in Figure \ref{fig:res3}. Again, both distributions agree within their uncertainties. While the observed mean of infections per 10k inhabitants is $261\pm9$, the simulated mean is $256\pm 9$. Even more interestingly, the observed RMS value, as measure for the spread, is $54\pm6$, which agrees well with the simulated RMS is $57\pm7$.

The agent based model \JuneGermany is therefore capable to describe the spread of infection numbers on local level, only using basic population statistics as well as common state-wide regulations without incooperating any specific local measures, such as information campaigns by the local authorities.

\section{Summary}

In this paper we use the \JuneGermany framework to predict the spread of infections within the German state of Rhineland Palatinate within the second COVID-19 wave from October 2020 and February 2021. The simulation was tuned on the overall number of confirmed COVID-19 death cases during October 2020 and mid of December 2021, however, no specific tuning on district level has been performed. We observe a good description of the spread of incidences per 10k inhabitants in all 36 districts and cities of Rhineland Palatinate. Since no specific local measures, such as information- , special testing or tracing campaigns have been implemented in the simulation, we conclude that those measures can have only a small or limited effect on the course of the pandemic. In contrary, the observed incidence rate in a given district is determined primarily by statistical effects, population statistics and state-wide regulations.

\section*{Acknowledgement}

This work has been supported by the Johannes Gutenberg Startup Research Fund. Part of the simulations were conducted using the supercomputer Mogon II at Johannes Gutenberg University Mainz. The authors gratefully acknowledge the computing time granted on the supercomputer.



\bibliographystyle{myelsarticle-num}
\bibliography{./Bibliography}

\end{document}